\documentclass[aip,amsmath,amssymb,reprint]{revtex4-1}

\pdfoutput=1

\usepackage{graphicx}
\usepackage{dcolumn}
\usepackage{bm}

\usepackage[utf8]{inputenc}
\usepackage[T1]{fontenc}
\usepackage{mathptmx}

\begin{document}

\preprint{AIP/123-QED}

\title{Synchronous single-photon detection with self-resetting GHz-gated superconducting NbN nanowires}

\author{E. Knehr}
 \email{emanuel.knehr@kit.edu}
 \affiliation{ 
Institute of Micro- and Nanoelectronic Systems, Karlsruhe Institute of Technology (KIT), 76187 Karlsruhe, Germany}
\affiliation{Department of Quantum Detection, Leibniz Institute of Photonic Technology, 07745 Jena, Germany}
 
\author{A. Kuzmin}
\altaffiliation[Now at ]{Laboratory for Applications of Synchrotron Radiation (LAS) and Institute of Photonics and Quantum Electronics (IPQ), Karlsruhe Institute of Technology (KIT), 76131 Karlsruhe, Germany}
\affiliation{ 
Institute of Micro- and Nanoelectronic Systems, Karlsruhe Institute of Technology (KIT), 76187 Karlsruhe, Germany}

\author{S. Doerner}
\author{S. Wuensch}
\author{K. Ilin}
\affiliation{ 
Institute of Micro- and Nanoelectronic Systems, Karlsruhe Institute of Technology (KIT), 76187 Karlsruhe, Germany}

\author{H. Schmidt}
\affiliation{Department of Quantum Detection, Leibniz Institute of Photonic Technology, 07745 Jena, Germany}
\affiliation{Institute for Solid State Physics, Friedrich Schiller University Jena, 07743 Jena, Germany}

\author{M. Siegel}
\affiliation{ 
Institute of Micro- and Nanoelectronic Systems, Karlsruhe Institute of Technology (KIT), 76187 Karlsruhe, Germany}

\date{14 September 2020}

\begin{abstract}
We demonstrate a GHz-gated operation of resonator-coupled superconducting nanowire single-photon detectors suitable for synchronous applications.
In comparison with conventional dc-biased nanowire detectors, this method prevents the detector from latching and can suppress dark counts and background noise.
Using a gating frequency of $3.8\text{~}\mathrm{GHz}$ and a fast, synchronized laser diode, we show that the detector's operation point follows the oscillating current and its detection efficiency depends on the relative frequency and phase of the bias and modulated optical signal. The obtained experimental results are in good agreement with simulations, showing that the duty cycle of a gated detector can be adjusted in a wide range in case of a pronounced saturation of the current-dependent detection efficiency.
This operation mode could be suitable for applications such as quantum key distribution and time-of-flight laser ranging.
\end{abstract}

\maketitle



Superconducting nanowire single-photon detectors (SNSPDs) offer many advantages over competing single-photon detector technologies. Due to their outstanding performance in terms of detection efficiency up to mid-infrared,\cite{Marsili:2012,Marsili:2013} detection speed,\cite{Kerman:2006a,Ferrari:2019} and timing precision,\cite{Korzh:2020} they are the single-photon detector of choice for many applications such as long-range optical communication,\cite{Shaw:2017} laser ranging,\cite{McCarthy:2013,Li:2016} and spectroscopy.\cite{Chen:2018}

SNSPDs are usually operated in free-running mode by applying a dc bias current slightly below the switching current of the nanowire. This is in contrast to single-photon avalanche diodes (SPADs), which require a gated operation to mitigate afterpulsing.\cite{Zhang:2015}
However, in applications with known photon arrival time spans, a gated operation is beneficial since background noise can be suppressed and the signal-to-noise ratio (SNR) therefore improved.
Additionally, gating the SNSPD using bias modulation can be used to reset the detector and prevent latching,\cite{Kerman:2009} which allows for a stable operation.

A gated or quasi-gated approach for SNSPDs has been adopted by various groups.
Akhlaghi \textit{et al.} presented a gated mode at $625\text{~}\mathrm{MHz}$ with a dc offset to increase the maximum count rate.\cite{Akhlaghi:2012}
To prevent latching due to unwanted backscattering in laser ranging (LIDAR), Zhang \textit{et al.} showed a dc operation with periodical off-states using a programmed bias source.\cite{Zhang:2017a}
For the application in quantum key distribution (QKD) systems, an auto-reset system for SNSPDs using a voltage-comparator was shown.\cite{Elezov:2019} This way, blinding attacks on the detector could be identified and mitigated.

A gated operation of SNSPDs using an rf bias in GHz range would push the operation closer to free-running mode, but requires considering the instantaneous bias current and its effect on the intrinsic detection efficiency (IDE) of the detector.
For nanowires made of superconductors with characteristic energy relaxation times much shorter than the period of the rf bias (e.g. NbN and $1\text{--}10\text{~}\mathrm{GHz}$ bias frequency), the IDE is expected to oscillate with twice the rf bias frequency, because it does not depend on the direction of the bias current.
In this case, the temporal distance between the active states of the detector is in the range of timing jitter of practical SNSPD devices (less than $100\text{~}\mathrm{ps}$ for an operation at $5\text{~}\mathrm{GHz}$).\cite{Kozorezov:2017}
This time interval could be further reduced if the IDE is well saturated over the bias current (i.e. it has a pronounced plateau in the sigmoidal count rate dependence $\mathrm{CR}(I_\mathrm{b}) \propto \mathrm{IDE}$). It should then also be possible to adjust the duty cycle of the gated operation in a broad range by changing the current amplitude.
Consequently, the operation of an SNSPD with a saturated $\mathrm{CR}(I_\mathrm{b})$-dependence could be freely adjusted between gated and nearly free-running mode according to the requirements of a particular application.

Previously, we demonstrated rf-biased SNSPDs which are embedded in a resonant circuit for frequency-division multiplexing of detector arrays (RF-SNSPD).\cite{Doerner:2016,Doerner:2017a} First evidence of a synchronous detection with these detectors was shown using a modulated light source at half of the gating frequency.\cite{Doerner:2019} 
Here, we present an RF-SNSPD with $3.8\text{~}\mathrm{GHz}$ gating frequency fully synchronized to a fast laser diode and demonstrate a pronounced synchronous single-photon detection, which is well described by the $\mathrm{IDE}(I_\mathrm{b})$-dependence.
For applications in which a synchronous detection can be implemented, the rf-biased, GHz-gated SNSPDs shown here could offer three main advantages: a reduction of dark counts, the mitigation of background noise, and a self-resetting, latching-free operation.



When a photon is absorbed by a dc-biased nanowire, a normal-conducting domain (hotspot) can form and its resistance manifests itself as a voltage pulse. If the nanowire's kinetic inductance is high enough (the nanowire is sufficiently long), the superconductive state can be restored and the detector is reset.
Latching into a normal state can occur if the nanowire is too short. In this case, a stable negative electrothermal feedback by the load impedance keeps the induced hotspot inside the nanowire in a self-heating state.\cite{Kerman:2009} The hotspot's lifetime strongly depends on the applied bias current.\cite{Kozorezov:2015,Marsili:2016} For currents below the retrapping current, the hotspot rapidly collapses.

In case of an RF-SNSPD, the nanowire is biased by a microwave current $I_\mathrm{rf}$ at the resonant frequency $f_\mathrm{r}$ of the nesting lumped-element resonator, which is capacitively coupled to a $50\text{~}\Omega$ feed line.
After photon absorption, a growing hotspot damps the resonator and therefore leads to an increased transmission coefficient $|S_{21}|$ on the feed line at $f_\mathrm{r}$, which is used as the detector response.

With this scheme, self-resetting of the nanowire is enforced by the oscillating current and the decay time of the resonator $\tau_\mathrm{d}$.
The hotspot growth time is limited to less than half a period of the rf current, since the hotspot collapses at low currents on a timescale of ${\sim}10\text{~}\mathrm{ps}$.
Additionally, the damping of the resonant circuit upon hotspot generation causes an rf current redirection out of the nanowire to the feed line for at least several periods. This effectively resets the detector after every detection event.

The self-resetting behavior also leads to a stable operation of the RF-SNSPD at bias amplitudes close to the switching current. With increasing rf bias, the dark count rate (DCR) reaches the maximum value $\mathrm{DCR} \approx 1 / \tau_\mathrm{d}$ and further increase of the current amplitude in the nanowire is limited. 
Instead, the RF-SNSPD demonstrates a self-oscillatory behavior, seen as strictly periodic ``dark counts'' with a rate linearly proportional to the ratio of the bias amplitude to the switching current ($\mathrm{DCR} \propto I_\mathrm{rf} / I_\mathrm{sw}$). If any excess optical power heats the nanowire, $I_\mathrm{sw}$ is reduced and the apparent DCR increases correspondingly.
This way, the RF-SNSPD with $\mathrm{DCR} > 1 / \tau_\mathrm{d}$ operates as a bolometric detector with ``power-to-frequency'' responsivity, similarly to a superconducting noise bolometer.\cite{Kuzmin:2017}

The decay time $\tau_\mathrm{d}$ of the RF-SNSPD resonator and, thus, the maximum count rate in single-photon regime can be designed to fit the application. Previously, an RF-SNSPD array with a decay time of only few ns was demonstrated,\cite{Doerner:2016} which is comparable to the reset time of typical dc-biased SNSPDs.

Furthermore, the rf bias generated by frequency synthesizers can be significantly more stable and is easier to filter than dc bias. This helps to avoid excess dark counts due to noise in the bias.



However, due to the oscillating bias current and the sigmoidal $\mathrm{IDE}(I_\mathrm{b})$-dependence, the RF-SNSPD is not expected to be active all the time, i.e., in gated mode.
In case of random photon arrivals, this should lead to a reduced detection efficiency compared to a similar dc-biased SNSPD.
For known photon arrival gates, on the other hand, the detection efficiency can be optimized by synchronizing the RF-SNSPD to the base clock of the optical signal. As a result, the RF-SNSPD's detection efficiency should become comparable to a dc-biased SNSPD.
In Ref.~\onlinecite{Doerner:2019}, the direct comparison of a dc- and rf-biased SNSPD has demonstrated that a similar asynchronous IDE can be achieved, although the comparison requires an accurate determination of the rf current in the nanowire, which is difficult. Moreover, the IDE of an RF-SNSPD synchronized to a modulated light source at half of the gating frequency revealed a sensitivity to the relative phase between the rf bias and optical modulation, although the change in the IDE was not well understood.
Here, we present results for an RF-SNSPD fully synchronized to a modulated light source in GHz range.

The RF-SNSPD used herein is fabricated from a $5\text{~}\mathrm{nm}$ thick NbN layer, similar to those presented in Ref.~\onlinecite{Doerner:2017}, and is operated at its resonant frequency $f_\mathrm{r} = 1.9\text{~}\mathrm{GHz}$. Thus, the gating frequency is $f_\mathrm{g} = 2 f_\mathrm{r} = 3.8\text{~}\mathrm{GHz}$.

As a GHz-modulated light source, we use a vertical-cavity surface-emitting laser diode (VCSEL, Optek Technology OPV314YAT) emitting light at a wavelength of $850\text{~}\mathrm{nm}$ and allowing for modulation frequencies $f_\mathrm{opt} \lesssim 4.5\text{~}\mathrm{GHz}$.

\begin{figure}
	\centering
	\includegraphics[width=0.46\textwidth]{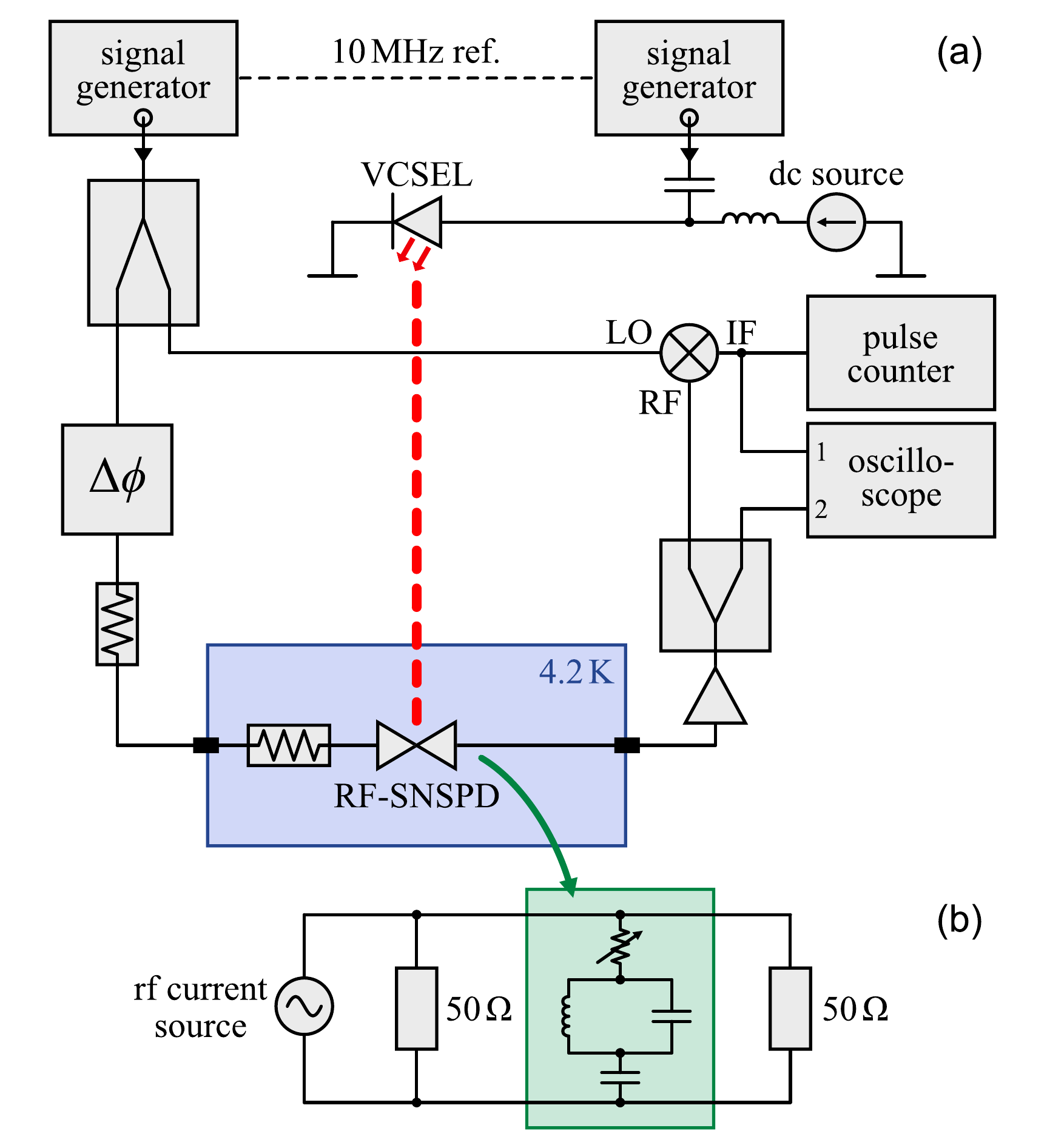}
	\caption{(a) Schematic of the experimental setup for synchronized photon detection. (b) Lumped-element circuit diagram of the resonator-coupled detector. The RF-SNSPD (green rectangle) consists of the photon-sensitive nanowire, depicted here as a variable resistor, a resonant circuit defined by an inductor and a capacitor, and a coupling capacitor.}
	\label{fig:setup}
\end{figure}

Two phase-locked rf generators provide the rf bias currents for both the laser diode and the RF-SNSPD (see Fig.~\ref{fig:setup}).
The laser diode is biased by a dc current of $3\text{~}\mathrm{mA}$ and an rf current with an amplitude large enough to turn it off at the minimum, resulting in an optical output power modulated in GHz range similar to on-off keying (OOK).
A fast photodiode (Thorlabs PDA8GS, $\mathrm{DC}\text{--}9.5\text{~}\mathrm{GHz}$) was used to accurately measure the spectrum of the optical modulation of the VCSEL.
The RF-SNSPD is biased by an rf current only.
The pulses from the detector are amplified by room-temperature amplifiers and read out by a $32\text{~}\mathrm{GHz}$ real-time oscilloscope.
Additionally, the rf pulses are downmixed by an analog mixer and the resulting envelope is read out with the osilloscope and an ordinary pulse counter to measure count rates.

\begin{figure}
	\centering
	\includegraphics[width=0.49\textwidth]{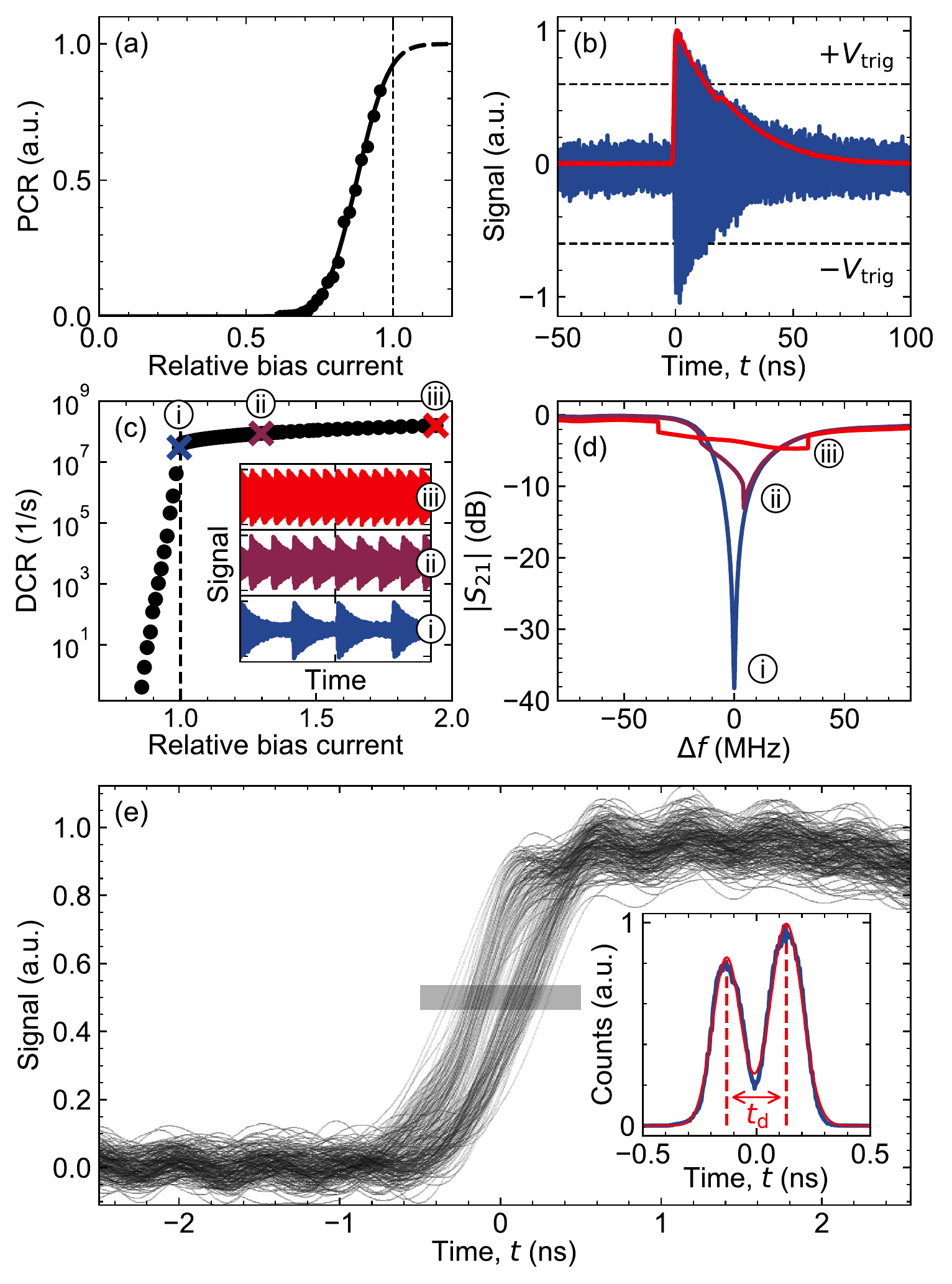}
	\caption{Current-dependent count rates and traces of detector pulses over time. (a) Measured photon count rate dependent on the relative bias current. The line is a theoretical fit. (b) Single rf (blue) and downmixed (red envelope) pulse upon photon absorption. (c) Dark count rate dependent on the relative bias current with insets depicting recorded waveforms at the highlighted bias levels. (d) Measured microwave transmission of the RF-SNSPD over frequency $\Delta f = f - f_\mathrm{r}$ at bias levels marked in (c). (e) Measured rising edge of multiple pulse envelopes. (Inset) Histogram of the normalized number of counts over time at the rising edge of the pulses (indicated by the semitransparent rectangle in (e)) with measurement in blue and Gaussian fits in red.}
	\label{fig:traces}
\end{figure}

The current-dependent photon count rate (PCR) of the detector is plotted in Fig.~\ref{fig:traces}(a), without saturation up to a level at which the DCR becomes dominant.

In Fig.~\ref{fig:traces}(b), a single detector response upon photon absorption and its envelope are shown.
The decay time $\tau_\mathrm{d} = Q_\mathrm{L}/\left( \pi f_\mathrm{r} \right) \approx 20\text{~}\mathrm{ns}$ is determined by the loaded quality factor $Q_\mathrm{L}$ of the resonant circuit.
The detector investigated herein is part of a frequency-multiplexed detector array, for which the value $Q_\mathrm{L} \approx 120$ was chosen to obtain a pixel frequency spacing in the range of several $10\text{~}\mathrm{MHz}$.

The latching-free operation of the RF-SNSPD is demonstrated by measuring the bias-dependent dark count rate (Fig.~\ref{fig:traces}(c)). Below a certain current amplitude (shown as 1.0), there is an exponential dependence of the DCR on the bias and dark counts are random. For relative bias currents above $1.0$, however, the DCR does not drop to zero but linearly rises and the pulses are periodic. We associate this operation with a bolometric regime as discussed before. For the highlighted bias currents, recorded rf pulses are depicted in the inset of Fig.~\ref{fig:traces}(c). In addition, microwave transmission measurements of the RF-SNSPD at the corresponding bias levels are shown in Fig.~\ref{fig:traces}(d). It can be seen that a cratered resonance appears above a certain bias level. However, the cratered resonance does not indicate that the nanowire is locked in the normal state but, in fact, periodically switches between superconducting and normal state.

Figure~\ref{fig:traces}(e) shows multiple traces of the envelope (channel~1 of the oscilloscope, Fig.~\ref{fig:setup}), which are obtained with constant optical power by triggering on the original rf pulse (channel~2) exceeding a defined range ($-V_\mathrm{trig}$ to $+V_\mathrm{trig}$, see Fig.~\ref{fig:traces}(b)). The polarity and amplitude of the first oscillation of the rf pulse depends on the phase of the bias current at the time of photon absorption. With this trigger method, one can visualize the distribution of count events relative to the phase of the bias.
Two distinct groups of envelopes can be observed.
This is also seen from the histogram of the number of counts, which is depicted in the inset of Fig.~\ref{fig:traces}(e). The time difference between the two peaks amounts to $t_\mathrm{d} \approx 263\text{~}\mathrm{ps}$, which corresponds to a frequency $f \approx 3.8\text{~}\mathrm{GHz}$ or twice the resonance frequency.

The grouping of pulses shown in Fig.~\ref{fig:traces}(e) can be explained by the non-saturated count rate characteristic $\mathrm{CR} \left( I_\mathrm{b} \right)$ of the NbN detector at $850\text{~}\mathrm{nm}$ and $4.2\text{~}\mathrm{K}$ (Fig.~\ref{fig:traces}(a)).
Since in this case the IDE does not reach $100\%$, the majority of the detection events will originate from bias points near the extrema ($\pm I_\mathrm{rf,max}$). 
The measured grouping of pulses around these two bias points of opposite polarity implies that the detector is active twice in a period of the rf bias. The gate duration can be estimated by the time during which $\left\vert I_\mathrm{rf} \right\vert \geq I_\mathrm{det}$, with the minimal detection current $I_\mathrm{det}$ for a particular photon energy.

In order to show that the RF-SNSPD's detection efficiency might be optimized by synchronizing the detector to the optical signal, the phase-sensitive changes of the count rate are investigated.
The measurements are conducted with the light modulated either with the resonance frequency of the RF-SNSPD ($f_\mathrm{opt} / f_\mathrm{r} = 1$) or with its gating frequency ($f_\mathrm{opt} / f_\mathrm{r} = 2$).
To slowly sweep the relative phase between the rf bias and optical signal, the frequency of the rf bias is detuned with respect to the modulation frequency of the laser diode by $\Delta f = 10\text{~}\mathrm{mHz}$. 
This small detuning ($\Delta f / f_\mathrm{r} \approx 10^{-12}$) should lead to a slow periodic modulation of the count rate with the beat frequency.

\begin{figure*}
	\centering
	\includegraphics[width=0.98\textwidth]{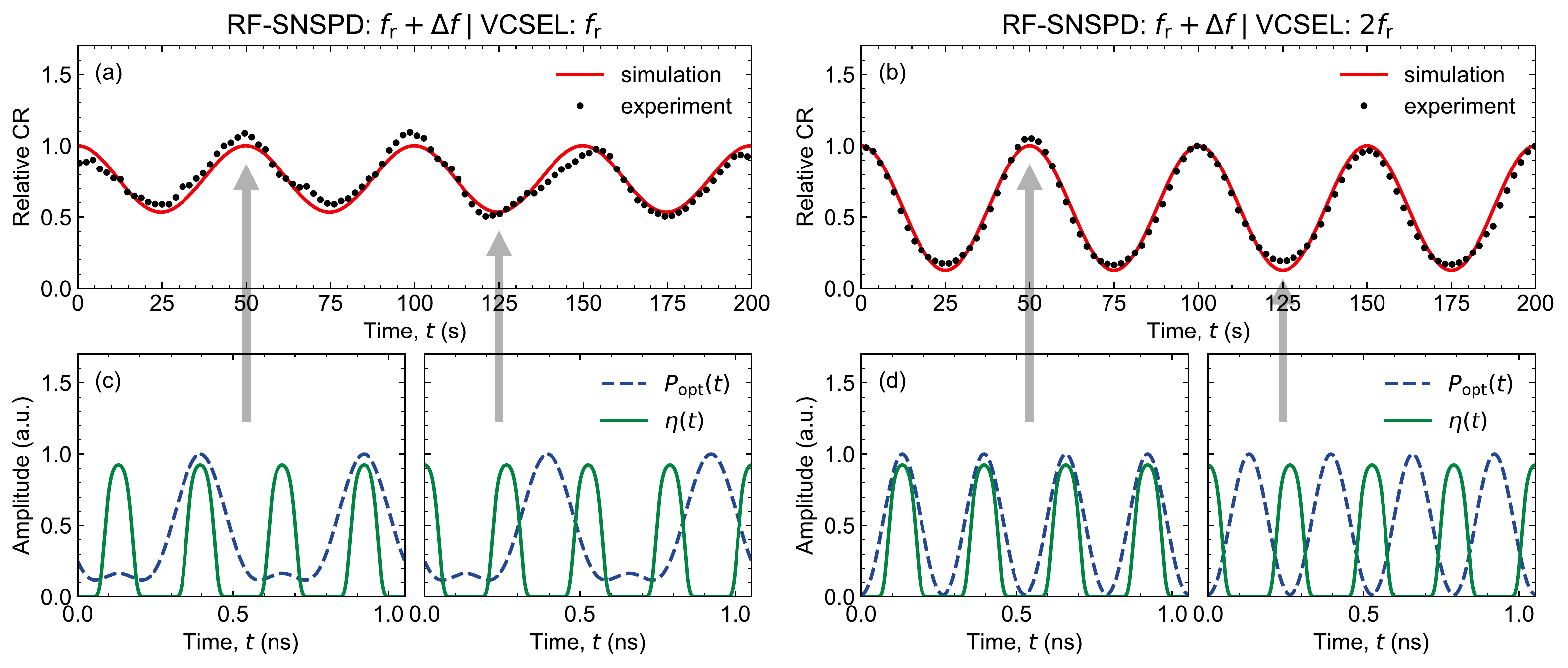}
	\caption{Measurement and simulation of modulated count rate $\mathrm{CR}(t)$ for a synchronized detection and beat frequency $\Delta f = 10\text{~}\mathrm{mHz}$. (a), (b) Experimental count rates $\mathrm{CR}(t)$ for $f_\mathrm{opt} / f_\mathrm{r} = 1$ and $f_\mathrm{opt} / f_\mathrm{r} = 2$ compared to simulated curves obtained from Eq.~(\ref{eq:CR_t}). (c), (d) Simulation of optical signal $P_\mathrm{opt}(t)$ and instantaneous detection efficiency $\eta(t)$ for both modulation frequencies $f_\mathrm{opt}$. In each case, full alignment and misalignment between $P_\mathrm{opt}(t)$ and $\eta(t)$ are depicted (in (d) corresponding to phases $0^{\circ}$ and $180^{\circ}$ of $\eta(t)$ relative to $P_\mathrm{opt}(t)$).}
	\label{fig:modulated_cr}
\end{figure*}

The phase-sensitive detection efficiency was analyzed experimentally (Fig.~\ref{fig:modulated_cr}(a) and (b)) and oscillations of the count rate with a period of $50\text{~}\mathrm{s}$ are indeed visible.
The measured count rate modulation index $m = (\mathrm{CR}_\mathrm{max} - \mathrm{CR}_\mathrm{min}) / (\mathrm{CR}_\mathrm{max} + \mathrm{CR}_\mathrm{min})$ amounts to $m \approx 0.71$ for $f_\mathrm{opt} / f_\mathrm{r} = 2$. For the case $f_\mathrm{opt} / f_\mathrm{r} = 1$, the modulation index is smaller ($m \approx 0.30$) and depends on the modulation amplitude of the laser diode.
The modulation period of $50\text{~}\mathrm{s}$ observed for a detuning $\Delta f = 10\text{~}\mathrm{mHz}$ corresponds to a detuning $\Delta f_\mathrm{g} = 20\text{~}\mathrm{mHz}$ in gating frequency, which is in accordance with the result in Fig.~\ref{fig:traces}(e).

To describe the measurement results, we simulate the $\mathrm{CR}(t)$-dependence using the instantaneous detection efficiency $\eta(t)$ obtained from the fit of $\mathrm{CR}(I_\mathrm{b})$.
For this, we take a sinusoidal rf bias current
\begin{equation}
	I_\mathrm{rf}(t) = \mathrm{sin} \left[ \left( \omega_\mathrm{r} + \Delta \omega \right) t \right],
\end{equation}
with $\omega_\mathrm{r} = 2 \pi f_\mathrm{r} = 2 \pi \times 1.9\text{~}\mathrm{GHz}$ and $\Delta \omega = 2 \pi \Delta f = 2 \pi \times 10\text{~}\mathrm{mHz}$.
The instantaneous detection efficiency $\eta(t)$ is then obtained by fitting
\begin{equation}
	\eta(t) = a \left[ 1 + \mathrm{erf} \left( b \left| I_\mathrm{rf}(t) \right| - c \right) \right]
	\label{eq:eta}
\end{equation}
to the experimental data (Fig.~\ref{fig:traces}(a), assuming $I_\mathrm{b} = I_\mathrm{rf}(t)$) with fitting parameters $a$, $b$, and $c$.
Since $\eta(t)$ in Eq.~(\ref{eq:eta}) depends on the modulus of the rf bias $|I_\mathrm{rf}| \propto \sqrt{1 - \mathrm{cos} \left( 2 \omega_\mathrm{r} t \right)}$, it oscillates with the frequency $f_\mathrm{g} = 2 f_\mathrm{r}$.

The amplitude modulated optical signal is modelled as
\begin{equation}
	P_\mathrm{opt}(t) = a_0 + a_1 \mathrm{cos} \left( \omega_\mathrm{opt} t  \right) + a_2 \mathrm{cos} \left(2 \omega_\mathrm{opt} t  \right),
\end{equation}
with the modulation frequency of the laser diode $\omega_\mathrm{opt} = 2 \pi f_\mathrm{opt}$ and experimentally measured spectral coefficients $a_0$, $a_1$, and $a_2$ for our particular laser diode at the used operation point.
Both the instantaneous detection efficiency $\eta(t)$ and the normalized optical power $P_\mathrm{opt}(t)$ are plotted over time in Fig.~\ref{fig:modulated_cr}(c) and (d) for operation frequencies $f_\mathrm{opt} / f_\mathrm{r} = 1$ and $f_\mathrm{opt} / f_\mathrm{r} = 2$, respectively.

The time-dependent count rate is estimated using Mandel's formula (Eq.~(8) in Ref.~\onlinecite{Mandel:1959}) by
\begin{equation}
	\mathrm{CR}(t) \propto \int_{t}^{t + \tau} \eta(t') \cdot P_\mathrm{opt}(t') \mathrm{d}t'
	\label{eq:CR_t}
\end{equation}
for the mean number of photon clicks in time interval $\tau$.

A comparison of the simulated and measured count rate $\mathrm{CR}(t)$ is shown in Fig.~\ref{fig:modulated_cr}(a) and (b).
As indicated by the arrows, the maxima and minima of $\mathrm{CR}(t)$ correspond to detector gates fully aligned and misaligned to the optical pulses, depicted in Fig.~\ref{fig:modulated_cr}(c) and (d).

For $f_\mathrm{opt} / f_\mathrm{r} = 2$, in which case the detector and the laser diode are operated with the same frequency and phase, the measurement data coincide well with the simulated curve both in beat frequency and modulation index $m$.
In this case, the amplitude $a_2$ of the second harmonic of $P_\mathrm{opt}(t)$ was found to be very small.
The large modulation index signifies that the detection efficiency can be maximized by synchronizing the detector to the light source.
The attenuation of stray light and suppression of dark counts can be quantified by the factor $\left( 1 + m \right) \cdot \left( 1 - m \right)^{-1}$, with a value of about $6.0$ for $f_\mathrm{opt} / f_\mathrm{r} = 2$.

For $f_\mathrm{opt} / f_\mathrm{r} = 1$, the modulation of $\mathrm{CR}(t)$ is only due to the second harmonic of $P_\mathrm{opt}(t)$, which has a significant amplitude $a_2$ originating from the nonlinear current-voltage characteristic of the laser diode. The latter also explains the observed dependence of modulation index $m$ on the bias amplitude of the diode.

\begin{figure*}
	\centering
	\includegraphics[width=0.98\textwidth]{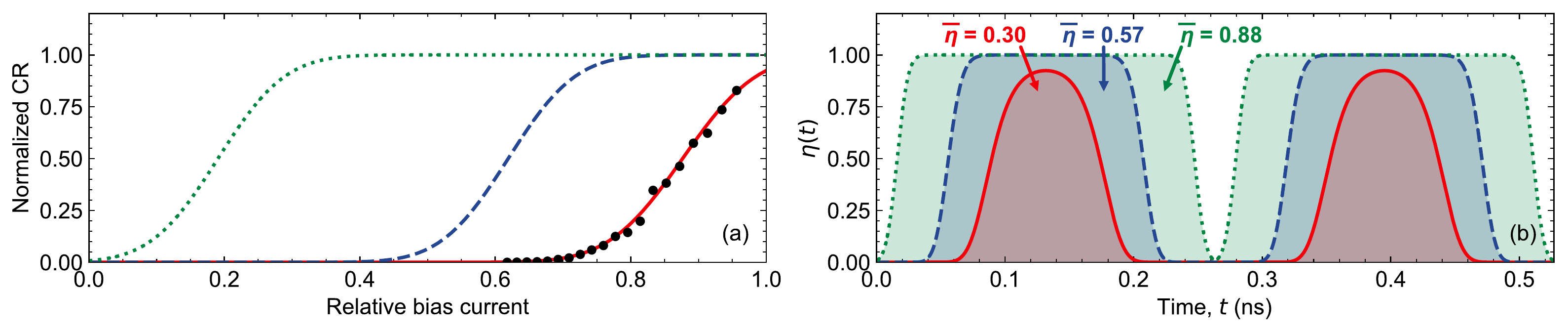}
	\caption{Simulation of different $\mathrm{CR}(I_\mathrm{b})$-curves (a) and corresponding instantaneous detection efficiencies $\eta(t)$ according to Eq.~(\ref{eq:eta}) over one period of $f_\mathrm{r}$ (b). The black dots in (a) indicate measurement results. Average detection efficiencies $\overline{\eta}$ in (b) are calculated from the integral of $\eta(t)$.}
	\label{fig:sim_saturated_CR}
\end{figure*}

According to our simple model, the duty cycle of the gated detector is highly dependent on the $\mathrm{CR}(I_\mathrm{b})$-characteristic.
With a non-saturated $\mathrm{CR}(I_\mathrm{b})$-curve, the investigated detector exhibits a gated detection with photon detection mainly occurring for a current $I_\mathrm{rf}$ near the extrema $\pm I_\mathrm{rf,max}$, as discussed before (see solid, red curves in Fig.~\ref{fig:sim_saturated_CR}).
In case of a pronounced count rate plateau, however, the active time spans can be much longer, which becomes important for detecting photons with random arrival times.
This is simulated for two saturated $\mathrm{CR}(I_\mathrm{b})$-curves (see dashed and dotted curves in Fig.~\ref{fig:sim_saturated_CR}). For a broader saturation in Fig.~\ref{fig:sim_saturated_CR}(a), the gate duration increases accordingly, as seen by the instantaneous detection efficiencies $\eta (t)$ in Fig.~\ref{fig:sim_saturated_CR}(b).
The average detection efficiency $\overline{\eta}$ can be estimated from the integral of $\eta (t)$, which results in values of $0.30$, $0.57$, and $0.88$ compared to a dc-biased SNSPD constantly operated at $\overline{\eta} = 1$.
For the broadest simulated $\mathrm{CR}(I_\mathrm{b})$-curve, the pure rf bias leads to an only slightly decreased efficiency for detecting random photon arrivals and therefore becomes comparable to free-running mode.
Thus, the duty cycle of an RF-SNSPD with a pronounced count rate saturation can be tuned between gated and quasi-free-running mode by choosing an appropriate rf bias amplitude.



In conclusion, we have shown a latching-free GHz-gated operation of resonator-coupled SNSPDs, which can be used for synchronous single-photon detection.
The detector is biased by a microwave current and its operation point is modulated accordingly.
Synchronous measurements indicate an instantaneous intrinsic detection efficiency following the rf bias current, which agrees well with our simple model of the phase-sensitive count rate. That way, we showed that the detection efficiency of RF-SNSPDs can be maximized and the influence of stray light minimized by using a phase-locked light modulated with a frequency twice the rf bias frequency.
The duty cycle of the detector can be much larger and tuned over a broad range for devices demonstrating a pronounced plateau in the bias dependence of the count rate. In this case, the detection efficiency of an RF-SNSPD is not degraded significantly even for random photon arrivals, which sets it on par with dc-biased, free-running SNSPDs.
Moreover, using a frequency-multiplexed detector array of RF-SNSPDs with moderate Q factors,\cite{Doerner:2016} a gated single-photon detection with a combined counting speed in GHz range could be possible.


\appendix

\section*{Data Availability}
The data that support the findings of this study are available from the corresponding author upon reasonable request.

\end{document}